\documentclass
[preprint,english,aps,pra,superscriptaddress,nofootinbib]{revtex4}%
\usepackage{amsfonts}
\usepackage[T1]{fontenc}
\usepackage{amsmath}
\usepackage{amssymb}
\usepackage{babel}
\usepackage{graphicx}
\usepackage{xcolor}%

\begin{document}
\title{Effect of a moving mirror on the free fall of a quantum particle in a
homogeneous gravitational field}
\author{J.\ Allam}
\affiliation{Laboratoire de Physique Th\'eorique et Mod\'elisation, CNRS Unit\'e 8089, CY
Cergy Paris Universit\'e, 95302 Cergy-Pontoise cedex, France}
\author{A.\ Matzkin}
\affiliation{Laboratoire de Physique Th\'eorique et Mod\'elisation, CNRS Unit\'e 8089, CY
Cergy Paris Universit\'e, 95302 Cergy-Pontoise cedex, France}

\begin{abstract}
We investigate the effect of time-dependent boundary conditions on the
dynamics of a quantum bouncer -- a particle falling in a homogeneous gravitational field on
a moving mirror. We examine more particularly the way a moving mirror
modifies the properties of the entire wavefunction of a falling particle. We
find that some effects, such as the fact that a quantum particle hitting a moving mirror may bounce significantly
higher than when the mirror is fixed, are in line with classical intuition. Other effects, such as the change in 
relative phases or in the current density in spatial regions arbitrarily far from the mirror are specifically quantum. We further discuss how
the effects produced by a moving mirror could be observed in link with current experiments, in
particular with cold neutrons.

\end{abstract}
\maketitle

\section{Introduction}

The quantum bouncer -- a quantum particle falling in a uniform gravitational
field bounded by a perfectly reflecting mirror -- is one of the paradigmatic
examples of quantum mechanics. Mentioned in some textbooks \cite{ref1,ref2},
the quantum bouncer has been used as a model to investigate wavepacket
dynamics and the quantum-classical correspondence \cite{gea,robinett}, to
derive semiclassical propagators \cite{ajp-path,jpa2021}, and, on the
experimental side, to identify the quantized eigenstates of cold neutrons
falling on a mirror in the Earth's gravitational field
\cite{exp-nature,rev2022}.

In this work, we will be interested in the dynamics of a quantum particle
obeying the Schr\"{o}dinger equation with a linear potential (due to a homogeneous
gravitational field) falling on a moving mirror.\ Such a problem belongs to
the class of systems subjected to time-dependent boundary conditions. Quantum
systems with time-dependent boundary conditions are interesting from a
mathematical \cite{mosta,martino}, foundational \cite{greenberger,A2018} or
practical \cite{pra-trans,epjd} perspective. Analytical solutions are known
only for some special systems \cite{mako1}.\ Even the simplest case -- an
infinite well with a moving wall -- needs to be solved numerically
\cite{glasser,mmw18}. Concerning experiments, setups with neutrons falling on
a moving mirror have been implemented in order to develop a gravity resonance
spectroscopy technique \cite{abele-spec} and test exotic theories of gravity
\cite{abele-np}.

Our aim here will be to investigate some effects induced by moving boundary
conditions on the dynamics of a quantum bouncer. Indeed, although
boundary conditions change the Hamiltonian only in a small spatial region, the
quantum-mechanical wavefunction changes everywhere, not only in the
neighborhood of that small
region. This implies that measurable effects (like the current density or
the difference in relative phases at a given point) due to moving boundaries can in principle be observed
everywhere as long as the wavefunction does not vanish. Such effects were
recently investigated in cavities with a moving wall \cite{mmw18} or in
confined time-dependent oscillators \cite{A2018,qr2020}. The original
motivation from a fundamental point of view was to look for a novel type of
single-particle non-locality \cite{greenberger}.\ In this paper we will be
extending these studies to a particle in free-fall bouncing on a mirror.

To this end, we will first briefly recall the basic features of the
Schr\"{o}dinger equation with time-dependent boundary conditions (Sec.\ \ref{freebc}) as well
as the main issues that appear when dealing with free fall and moving
boundaries (Sec.\ 3). We investigate in Sec.\ 4 the evolution of quantum
properties in the presence of moving boundaries.\ This will be done by
comparing the dynamical evolution of a given initial state in the presence of
fixed and moving boundaries. We will compare the short-time as well as the
long-time dynamics for different types of initial states.\ We will particularly focus on the
evolution of the current density, the phase,and the probability density of a bouncing wavepacket. We will discuss our results in
Sec.\ \ref{sec5}, in particular on the prospects for observing
experimentally the effects investigated in this work. A short Conclusion is provided in Sec. \ref{conc}.

\section{Schr\"{o}dinger equation with time-dependent boundary
conditions\label{freebc}}

Before getting to the problem of a quantum particle bouncing on a moving
mirror (Sec.\ \ref{moving}), let us briefly look at the simplest system with
time-dependent boundary conditions: a quantum particle placed in an infinite
well with one of the walls (say the right edge) moving according to some
function $L(t)$.\ Such a system is defined (see eg \cite{physica}) by the
Hamiltonian
\begin{align}
H  &  =\frac{P^{2}}{2m}+V\label{ham}\\
V(x)  &  =\left\{
\begin{array}
[c]{l}%
0\text{ \ for}\ \ 0\leq x\leq L(t)\\
+\infty\text{ \ otherwise.}%
\end{array}
\right.  \label{vdef}%
\end{align}
where $m$ is the mass of the particle and $L(t)$. The solutions of the
Schr\"{o}dinger equation must obey the boundary conditions $\Psi(0,t)=\Psi
(L(t),t)=0$, so that formally a different Hilbert space needs to be defined at
each $t$ \cite{martino}. In such problems it is important to distinguish the
instantaneous eigenstates of $H$,
\begin{equation}
\phi_{n}(x,t)=\sqrt{2/L(t)}\sin\left[  n\pi x/L(t)\right]  \label{eig}%
\end{equation}
that verify
\begin{equation}
H\left\vert \phi_{n}(t)\right\rangle =E_{n}(t)\left\vert \phi_{n}%
(t)\right\rangle
\end{equation}
(where $E_{n}(t)=n^{2}\hbar^{2}\pi^{2}/2mL^{2}(t)$ are the instantaneous
eigenvalues), from the function basis solutions of the Schr\"{o}dinger equation%
\begin{equation}
i\hbar\partial_{t}\psi_{n}(x,t)=H\psi_{n}(x,t). \label{tdse}%
\end{equation}
Indeed, the eigenstates $\phi_{n}(x,t)$ do not obey the Schr\"{o}dinger equation. If the
evolution of an initial wavefunction, say $\Psi(x,0),$ is sought in terms of
linear superposition, then the expansion should be written in terms of the
basis functions as $\Psi(x,0)=\sum_{n}c_{n}\psi_{n}(x,0),$ since the
evolution will read%
\begin{equation}
\Psi(x,t)=\sum_{n}c_{n}\psi_{n}(x,t). \label{expm}%
\end{equation}

Unfortunately, the basis functions cannot be obtained in closed form except
for special choices of functions $L(t)$; for example for a linear expansion
$L(t)=L_{0}+at$ the functions $\psi_{n}(x,t)$ have a particularly simple form
\cite{doescher}. For an arbitrary bounday motion $L(t)$, numerical methods
need to be employed. One method is to look for an expansion
\begin{equation}
\sum_{n}c_{n}(t)\phi_{n}(x,t) \label{expa}%
\end{equation}
in the eigenstate basis with time-dependent coefficients $c_{n}(t),$ that are
obtained by solving numerically a system of coupled first order differential
equations \cite{glasser}. This method works efficiently if the coupled system
matrix elements decrease to zero fast enough as the basis size is increased;
this essentially depends on the scalar products $\left\langle \phi
_{n}(t)\right\vert \left.  \dot{\phi}_{m}(t) \right\rangle $ (the overdot here
and below labels the time-derivative). This is indeed the case \cite{mmw18}
for the basis states given by Eq.\ (\ref{eig}).

\section{Quantum bouncer}

\subsection{Fixed mirror\label{bouncer}}

The quantum bouncer problem is defined by the Schr\"{o}dinger equation of a
particle of mass $m$ falling on a mirror in a homogeneous gravitational
field,
\begin{equation}
ih\partial_{t}\Psi(z,t)=\left(  -\frac{\hbar^{2}}{2m}\partial_{z}%
^{2}+V(z)\right)  \Psi(z,t) \label{schrod}%
\end{equation}
where $V(z)=mgz$ is the gravitational potential ($g$ is the local free fall
acceleration). Placing the perfectly reflecting mirror at $z=0$ implies the
wavefunction must obey the boundary condition $\Psi(0,t)=0$. In order to
obtain the energy eigenvalues of the time-independent Schr\"{o}dinger equation it
is customary (e.g., Ref. \cite{gea}) to rescale the variables, $z\rightarrow
z/l_{g}$, $E\rightarrow E/(mgl_{g})$ yielding
\begin{equation}
\frac{d^{2}\phi_{n}}{dz^{2}}-(z-E_{n})\phi_{n}(z)=0,
\end{equation}
whose solutions are well known to be given in terms of the regular Airy
function $\mathrm{Ai}(z)$ by
\begin{equation}
\phi_{n}(z)=N_{n}\mathrm{Ai}(z-z_{n}). \label{airyfi}%
\end{equation}
The notation $z_{n}\equiv E_{n}$ is chosen because these are simply related to
the zeros of the Airy function by $\mathrm{Ai}(-z_{n})=0$. $N_{n}$ is a
normalization constant that can be obtained in closed form \cite{robinett}.
The length in these units is given in terms of $l_g =(h^2/(2 g m^2))^{1/3}$, while time is given in terms of
$t_{g}=\left(  2\hslash/\left(  g^{2}m\right)  \right)  ^{1/3}.$

The quantized states $\phi_{n}(z)$ of a particle falling on a mirror have been
observed experimentally with ultracold neutrons \cite{exp-nature}. The
dynamics of an arbitrary initial state can be easily computed by expanding the
initial wavefunction in terms of $\phi_{n}(z),$ from which it follows that
$\Psi(z,t)=\sum_{n}\left\langle \phi_{n}\right\vert \left.  \Psi\right\rangle
e^{-iE_{n}t/\hbar}\phi_{n}(z)$. Several dynamical features of eigenstates or
wavepackets have been investigated theoretically \cite{gea,robinett}.

\begin{figure}[h]
	\centering
	\includegraphics[scale=0.2]{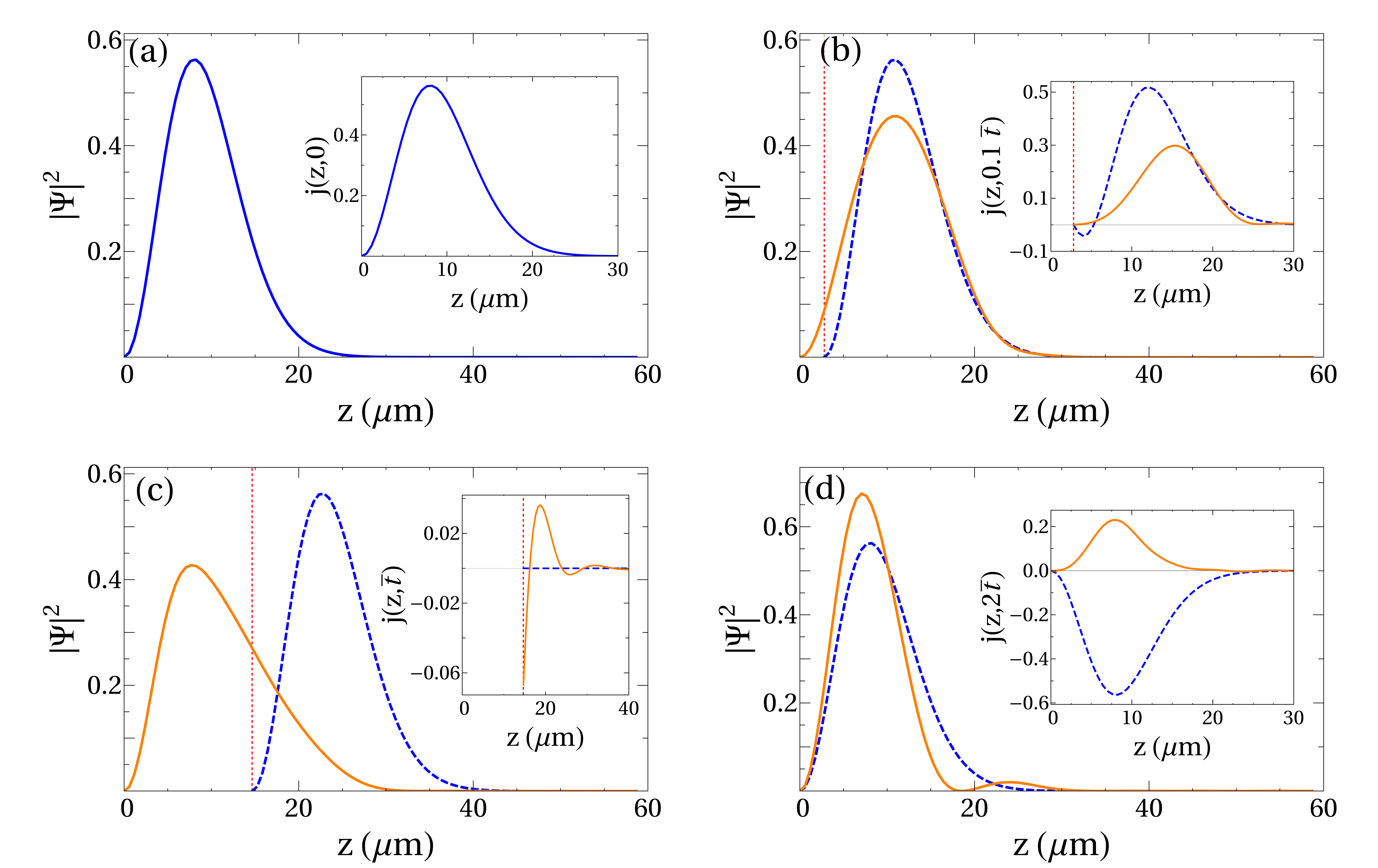}
	\caption{Snapshots of the time evolution of the probability density and the current density (shown in the insets)  for a neutron prepared in the ground basis state of the moving mirror $\Psi(z,0)=\psi_{1}(z,0)$. (a) shows the initial state ($t=0$); (b)-(d): The probability density $|\Psi^{F}|^2$ and corresponding current when the mirror is fixed are shown in orange; the blue dotted lines represent $|\Psi^{M}|^2$  and $j_{1}^{M}$ (inset) when the mirror is moving. The position of the mirror is indicated by a vertical red line. These snapshots are taken (b) at $t=\bar{t}/10$; (c) at $t=\bar{t}$ (when the mirror reaches its maximum position); (d) at $t=2\bar{t}$ (the mirror returns to its initial position). The parameters of the mirror in gravitational units are $ a=-0.1 $, $b=1$ and $c=0$, which sets $\bar{t}=5 t_{g}$. The values shown in the plots have been rescaled from the gravitational unit values by using the neutron mass and the Earth's surface gravitational acceleration $g$, giving $L(\bar{t})=14.67 \mu m$ at $\bar{t}=5.47 $ ms.  }
	\label{fig-mak}
\end{figure}

\subsection{Moving mirror\label{moving}}

We assume now the mirror is moving, with its position given by $L(t).$
Eq.\ (\ref{schrod}), as well as the Hamitonian (\ref{ham}) still hold with $V$
now given by
\begin{equation}
V(z)=%
\begin{cases}
+\infty & z< L(t)\\
mgz & z\geq L(t)
\end{cases}
. \label{vt}%
\end{equation}
This is similar to Eq. (\ref{vdef}) implying time-dependent boundary
conditions: a solution $\Psi(z,t)$ of the Schr\"{o}dinger equation must obey
$\Psi(z=L(t),t)=\Psi(z=\infty,t)=0.$ The instantaneous eigenstates analogous
to Eq. (\ref{eig}) are given here by
\begin{equation}
\phi_{n}(z,t)=N_{n}\mathrm{Ai}(z-z_{n}-L(t)), \label{inst}%
\end{equation}
but, as above, they are not solutions of the Schr\"{o}dinger equation
(\ref{schrod})-(\ref{vt}). We have found that their use in numerical schemes
is not as reliable as in the infinite well case mentioned in
Sec.\ \ref{freebc} due to the properties of the Airy functions: scalar
products of the type $\left\langle \phi_{n}(t)\right\vert \left.  \dot{\phi
}_{m}(t)\right\rangle $ decrease very slowly with increasing $m$, so that
hundreds of thousands of terms would be needed in the expansion (\ref{expa})
for typical parameter values.

Closed-form solutions of the Schr\"{o}dinger equation with the potential
(\ref{vt}) with an arbitrary function $L(t)$ are unknown except in a handful
of special cases \cite{mako1,mako2}. One particular case we will be using
below is the parabolic motion
\begin{equation}
L(t)=at^{2}+bt+c. \label{parabola}%
\end{equation}
\ The solutions can be shown \cite{mako1} to be given by
\begin{equation}
\psi_{n}(z,t)=\mathcal{N}_{n}e^{-\frac{im}{2\hslash}\left(  2\dot
{L}(t)(z-L(t))+\int_{0}^{t}\dot{L}(t^{\prime})^{2}dt\prime-2g\int_{0}%
^{t}L(t^{\prime})dt\prime\right)  }e^{-\frac{i}{\hslash}\Lambda_{n}t}Ai\left(
\frac{z-L(t)}{\delta}-z_{n}\right)  , \label{psi}%
\end{equation}
where%
\begin{equation}
\delta=\left[  \frac{\hslash^{2}}{2m^{2}(g+\ddot{L}\left(  t\right)
)}\right]  ^{1/3},
\end{equation}
$\ddot{L}=2a$ has to be such that $g+2a>0$, and $\Lambda_{n}$ is related to
the $n$th zero of the Airy function through
\begin{equation}
\Lambda_{n}=m\delta\left(  g+\ddot{L}\left(  t\right)  \right)  z_{n}.
\end{equation}
The normalization constant (not given in Ref. \cite{mako1}) can be found to be
given by%
\begin{equation}
\mathcal{N}_{n}=\left(  \sqrt{\delta}\mathrm{Ai}^{\prime}\left(  z_{n}\right)
\right)  ^{-1}.
\end{equation}
These functions $\psi_{n}$ form a time-dependent basis over which any initial
wavefunction can be expanded through Eq. (\ref{expm}).

As a consequence of applying the rescaling $z\rightarrow
z/l_{g}$, $E\rightarrow E/(mgl_{g})$, and $t\rightarrow
t/t_{g}$ (Sec. \ref{bouncer}), we get $g=2$ and $\hslash=2m$ when expressed in gravitational units. This will give 
$\delta=\left[  \frac{2}{2+\ddot{L}\left(  t\right)}\right]  ^{1/3}$, and  $\Lambda_{n}=2m z_{n}/\delta^2$. Consequently, $\psi_{n}(z,t)$ can be written in scaled units as 
\begin{equation}
\psi_{n}(z,t)=\mathcal{N}_{n}e^{-\frac{i}{4}\left(  2\dot
{L}(t)(z-L(t))+\int_{0}^{t}\dot{L}(t^{\prime})^{2}dt\prime-4\int_{0}%
^{t}L(t^{\prime})dt\prime\right)  }e^{-i z_{n} t/\delta^2}Ai\left(
\frac{z-L(t)}{\delta}-z_{n}\right)  . \label{psi}%
\end{equation}

\section{Results\label{res}}

We compare in this section the evolution of a quantum particle prepared in a
chosen initial state with a fixed or moving mirror.\ The initial state will be
set to be (i) a basis function of the moving mirror, (ii) an eigenstate of a
falling particle with a fixed mirror, or (iii) a Gaussian wavepacket prepared
above the mirror. In all cases considered we will assume the mirror moves
parabolically according to Eq. (\ref{parabola}).

\subsection{Initial basis state of the moving mirror
bouncer\label{init-moving}}

Let us choose as the initial state the ground basis state of the mirror with
parabolic motion, given by Eq. (\ref{psi}), $\Psi(z,0)=\psi_{1}(z,0)$.

The subsequent evolution if the mirror follows the parabolic motion is then%
\begin{equation}
\Psi^{M}(z,t)=\psi_{1}(z,t)
\end{equation}
and can be read off from Eq. (\ref{psi}): the modulus is an Airy function
shifted by the mirror's motion, while the exponential includes a phase as well as a
spatial dependent term. If $a$ and $b$ in Eq. (\ref{parabola}) have opposite
signs, there is a time $\bar{t}$ for which $\dot{L}(\bar{t})=0$ at which point
the spatial dependence becomes real.\ This implies the current density should
vanish at this time, as can be confirmed by computing
\begin{equation}
j_{n}^{M}(z,t)=\dot{L}(t)\left[  \mathrm{Ai}\left(  z_{n}-\frac{(c+t(b+at)-z)}%
{\delta}\right)  \right]  ^{2},
\end{equation}
and hence $j_{n}^{M}=0$ everywhere whenever $\dot{L}(t)=0$.  

If we prepare the system in the same initial state $\psi_{1}(z,0)$ but the
mirror remains fixed, the ensuing evolution is different.\ The wavefunction is
of course not shifted. By expanding $\psi_{1}(z,0)=\sum_{n}c_{n}\phi_{n}(z)$
in the basis of fixed mirror eigenstates (\ref{airyfi}), with $c_{n}%
=\left\langle \phi_{n}\right\vert \left.  \psi_{1}\right\rangle $, we obtain
the evolution in the fixed mirror case as
\begin{equation}
\Psi^{F}(z,t)=\sum_{n}c_{n}e^{-iE_{n}t/\hbar}\phi_{n}(z).
\end{equation}

Numerical results are shown in Fig. \ref{fig-mak}. The intial wavefunction
[Fig. \ref{fig-mak}(a)] evolves differently when the mirror is fixed or
moving.\ Even for very short times, when the mirror has barely moved, the
wavefunction and the current density are appreciably different [Fig.
\ref{fig-mak}(b)]. When the mirror reaches its maximum, the current density
$j_{n}^{M}$ vanishes everywhere in the moving case, but not in the fixed case [Fig.
\ref{fig-mak}(c)].
When the mirror comes back at its initial position (at $t=2\bar{t})$, we have
$\left\vert \Psi^{M}(z,2\bar{t})\right\vert ^{2}=\left\vert \Psi
^{M}(z,0)\right\vert ^{2}$ for the probability density when the mirror moves,
but this is seen not to be the case if the mirror remains fixed [Fig.
\ref{fig-mak}(d)]; note that the probability density in the fixed and moving mirror cases flows in opposite directions.

\begin{figure}[h]
	\centering
	\includegraphics[scale=0.2]{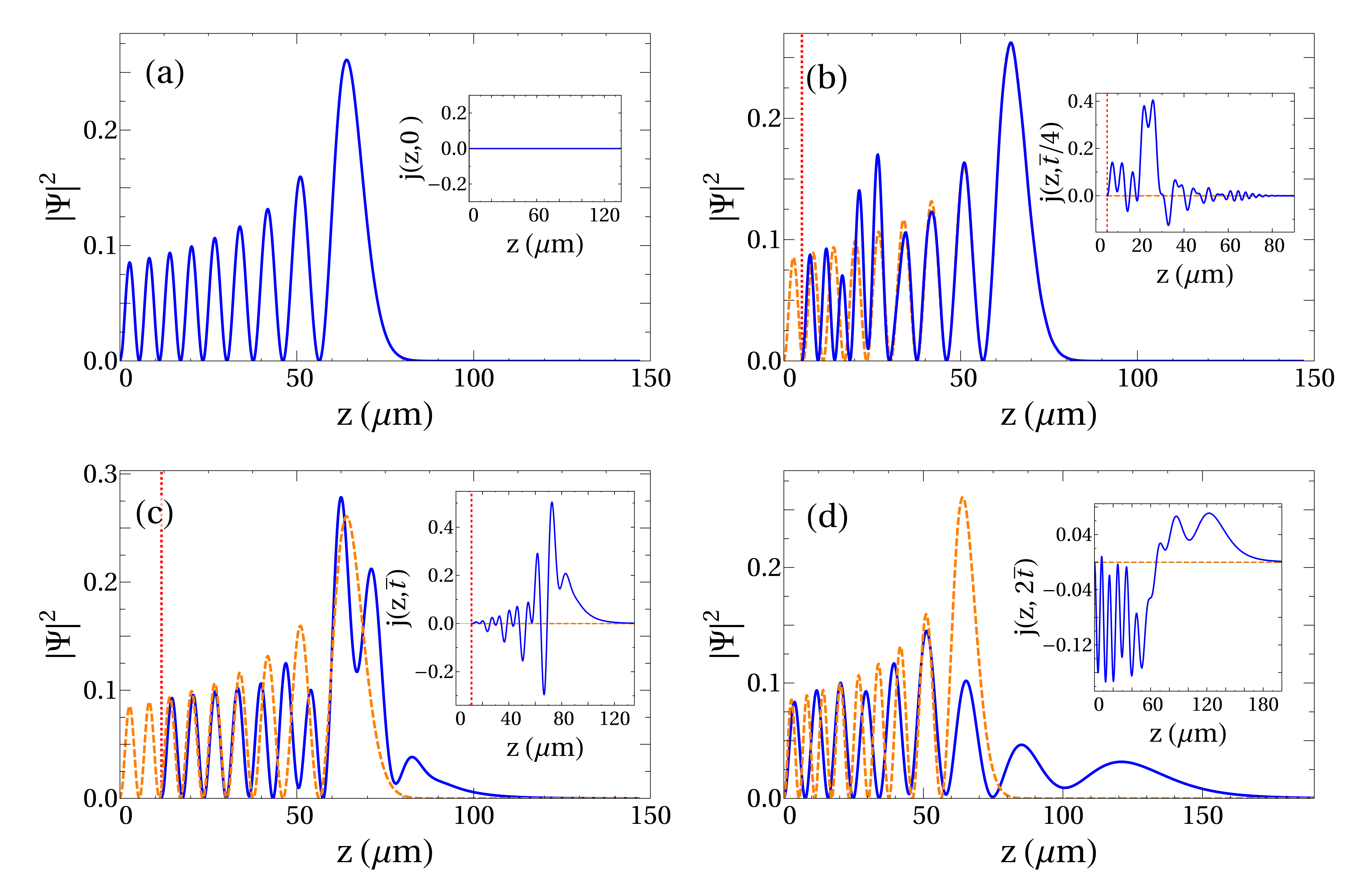}
	\caption{Similar to Fig. \ref{fig-mak} but when the initial state is prepared as the $n=9$ excited eigenstate of the \textit{fixed} mirror ($\Psi(z,0)=\phi_{9}(z)$) and for a different parabolic motion ($a=-0.5$,$b=2$,$c=0$ in scaled units). (a) shows the initial probability density. (b) shows the probability and current densities (in the insets) at $t=\bar{t}/4$ when the mirror is fixed (dashed orange line) or moving (solid blue). The vertical red line displays the position of the mirror in the moving case. (c): Same as (b) but for 
		$t=\bar{t}$, when the mirror is at its maximum position. (d) shows the same quantities at $t=2\bar{t}$ when the mirror returns to $z=0$. The values shown in the plots are obtained by rescaling the gravitational units to SI units for a neutron  falling in the gravitational field at the surface of the Earth, giving a maximum position of the mirror $L(\bar{t})=11.73 \mu m$ at $\bar{t}=2.18 ms$.  }
	\label{fig-fixed}
\end{figure}

\subsection{Initial eigenstate of the fixed mirror bouncer\label{init-fixed}}

We now choose as the initial state an excited state (say the $k$th excited
state) of the fixed mirror problem, so that $\Psi(z,0)=\phi_{k}(z)$ as given
by Eq. (\ref{airyfi}). If the mirror is fixed, the system remains in the
eigenstate and the evolution of $\Psi^{F}(z,t)$ is trivial.\ If the mirror
moves, we need to expand the initial state in terms of the basis functions
$\Psi(z,0)=\sum_{n}d_{n}\psi_{n}(z,0)$ with $d_{n}=\left\langle \psi
_{n}(t=0)\right\vert \left.  \phi_{k}\right\rangle $, and the evolution
becomes
\begin{equation}
\Psi^{M}(z,t)=\sum_{n}d_{n}\psi_{n}(z,t).
\end{equation}

An illustration is given in Fig. \ref{fig-fixed} for the $k=9$th excited
state. Even for short times, there is a significant difference between the
evolution in the fixed and moving mirror cases, as the nodal structure of the
moving case is pushed by the mirror at distances far beyond the range of
motion of the mirror. Although the highest position of the mirror is $z=11.73$ $\mu$m, the bouncing particle explores regions of space twice as high
than in the fixed mirror case, as basis states $\psi_{n}$ with very high $n$
are populated. Note that the current density, that remains zero everywhere 
when the mirror is fixed, varies wildly if the mirror moves, including for high values of $z$.

\subsection{Initial Gaussian wavepacket}

We finally consider the quantum particle is initially prepared as a Gaussian
wavepacket falling with zero average momentum towards the mirror. The initial
wavefunction%
\begin{equation}
\Psi(z,0)=\sqrt{\frac{2}{\pi  \sigma ^2}} \exp \left(-\frac{(x-\text{z0})^2}{\sigma ^2}\right)
\end{equation}
is expanded over the eigenstates of the fixed miror Hamitonian, in order to
compute $\Psi^{F}(z,t)$, or on the basis function of the Hamiltonian with the
moving boundaries, in order to compute $\Psi^{M}(z,t)$. 

\begin{figure}[htbp]
	\hspace*{-1cm}                                                            
	\includegraphics[scale=0.6]{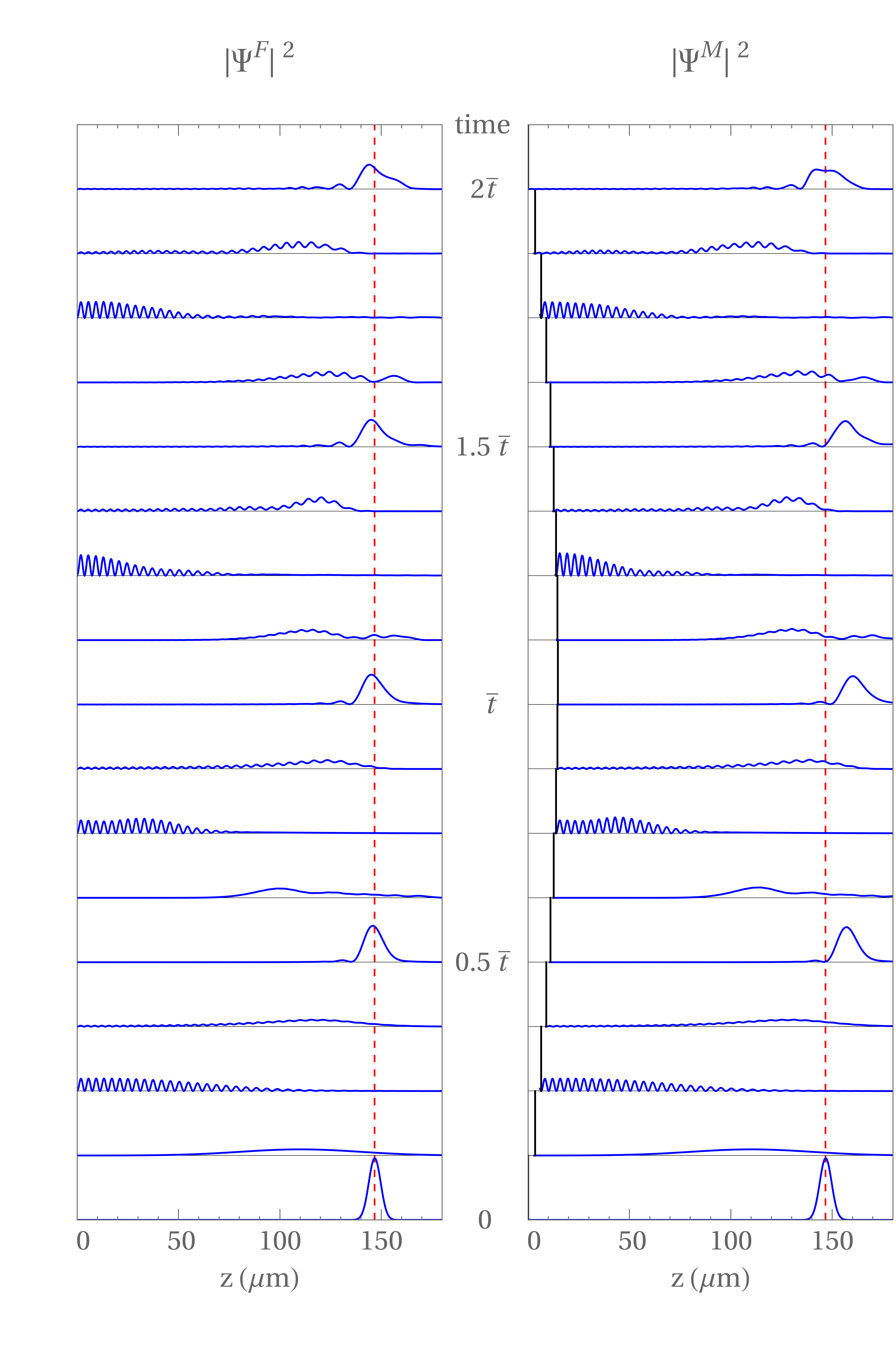}\caption{Evolution of an initial Gaussian wave packet released at rest from a height $z_{0}=25 l_{g}$, and bouncing over a fixed mirror (left panel) or over a moving mirror (right panel). The initial Gaussian is at the bottom ($t$ increases from bottom to top); the red dotted line shows the average position while the parabolic black curve on the right panel indicates the mirror's position (the parameters of the mirror are in  gravitational units $a=-0.00625$, $b = 0.25$, and $c=0$). The mirror returns to its initial position at $t=2\bar{t}$. We have rescaled the gravitational units so that the numbers on the axes correspond to a neutron falling in the gravitational field of the Earth, leading to $z_{0}=146.7 \mu m$, $\bar{t}=21.88 ms$ and $L(\bar{t})=14.67 \mu m$.}
	\label{fig-Gau}
\end{figure}

The results for a
typical example are illustrated in Fig. \ref{fig-Gau}, comparing the wavepacket dynamics for a particle initially in a Gaussian state falling on a fixed or moving mirror. In this case the mirror moves upwards, kicking the wavepacket to higher altitude than for the fixed mirror bouncer.

%\begin{figure}[htbp]
%	\hspace*{-2cm} 
%	\includegraphics[scale=0.8]{average.png}\caption{Expectation value $\langle z \rangle$ of the position  versus time for both cases displayed in figure (\ref{gauss}). The fixed mirror case is shown as a dashed line, and the moving mirror case as a solid line. The motion of the mirror is also shown as the parabolic curve below. The expectation value of the particle bouncing over the moving mirror shifts upwards along the motion of the mirror as expected classically.}
%\end{figure}

\section{Discussion}
\label{sec5}

\subsection{Dynamical effects}

We have seen for different types of initial states, that the dynamics of a
quantum bouncer falling on a moving mirror is different from the bouncer
evolution when the mirror is fixed. Some aspects of this behavior can be
understood intuitively on classical grounds, for example the fact that the
bounce is higher when the particle is kicked upwards by the moving mirror.
However there are specific quantum features.\ These are due to the existence
of discrete energy levels of the particle in the gravitational field, and to
the fact that the wave needs to adapt globally to changing boundary conditions.

As a result, the effect of the mirror's motion on the wavefunction and hence
on observable properties can be monitored at any point in space. For example
the moving mirror populates eigenstates of the gravitational field with higher
energies (see Sec. \ref{init-fixed}), a process that is at the basis of recent
proposals \cite{pra-trans,epjd} aiming to implement a spectroscopic method to
induce transitions between states. Another striking effect is that the current
density may vanish everywhere when the mirror reaches an extremum position,
when the initial state is prepared in a basis function state (see Sec.
\ref{init-moving}). In this situation, we also have $\left\vert \Psi
^{M}(z,2\bar{t})\right\vert ^{2}=\left\vert \Psi^{M}(z,0)\right\vert ^{2}$
when the mirror returns to its original position at time\ $t=2\bar{t}$, but
the argument of the wavefunction is almost opposite, i.e. $\arg\Psi
^{M}(z,2\bar{t})\simeq-\arg\Psi^{M}(z,0)$ as can be checked from Eq.
(\ref{psi}) (with $L(2\bar{t})=L(0)$ and $\dot{L}(2\bar{t})=-\dot{L}(0)$). The examples pictured
in Figs. \ref{fig-mak} and \ref{fig-fixed} clearly indicate that boundary conditions have an
effect on the entire wavefunction well beyond the region of changing boundaries.

\subsection{Non-locality}

The dynamical features produced by the moving mirror might give rise to
non-local effects. The idea that moving boundaries could induce non-locality
was proposed by Greenberger \cite{greenberger}.\ Although Greenberger's
original proposal (based on a wavepacket positioned far from the moving
boundary) was disproved \cite{A2018}, it was realized that a quantum state
having non-zero probability in the moving boundary region could display
non-local features in the sense that an observer located far from the moving
boundary region could guess whether the mirror is moving or not before a light
ray emitted from the moving boundary region reaches the observer's position.
Note that this instantaneous reaction, albeit small, of the wavefunction to
changing boundary condition does not seem to be an artifact of the
non-relativistic framework, as it was also seen to be the case for
relativistic wavefunctions \cite{colin20}.

In the cases examined in Sec. \ref{res}, non-locality, as defined here, would
be obtained provided the dynamical features due to the moving mirror can be
observed before a light signal emitted from the position of the moving
boundary reaches the point of observation. For instance, when a basis function
is chosen as the initial state (Sec. \ref{init-moving}), one would need to
observe the current density at point $z_{0}$, or check the relative phase
between the inital and time evolved quantum states at $t=2\bar{t}$ (by
employing an interferometric setup \cite{qr2020}), in a time shorter than
$z_{0}/c$. When the initial state is chosen to be an eigenstate $\phi_{k}(z),$
there are positions $\tilde{z}$ for which $\left\vert \phi_{k}(z)\right\vert
^{2}$ vanishes, while the time-evolved state $\Psi^{M}(z,t)$ in case the
mirror moves will have a small but non-vanishing probability to be detected
(see Fig. \ref{fig-fixed}); such detections could in principle take
place outside the light cone emanating from the mirror position.

\subsection{Experimental observation}

Effects produced by a moving mirror on a quantum particle in free fall in a
homogeneous gravitational field have actually already been observed
experimentally with ultra-cold neutrons \cite{abele-spec} when implementing a
resonance spectroscopy method. Other effects based on particle counting might
be observable with the same technology. The current density coud be observed
by coupling the neutron spin to the vertical velocity component with a
wedge-shaped magnetic field \cite{rauch}, linking the spin deflection to the
velocity. Quantities related to relative phases would require interferometry
based techniques, the most straightforward option (a neutron bouncing in a
Mach-Zehnder interferometer with different boundary conditions in each arm)
appearing as hardly realizable.

The observation of non-local aspects\footnote{Of course, one would actually
rather expect on physical grounds to see the breakdown of the wavefunction
formalism rather than detection of non-local aspects that could be signalling
in certain cases, see Ref. \cite{qr2020}. Still, a negative result would be
extremely valuable.} appear also difficult to implement with present day
technologies.  In principle, the detection of sufficient neutrons (for example
by simultaneously measuring on a high number of identical systems the number
of detections at a height larger than $\tilde{z}$) in regions for which the
fixed mirror case would lead to no detections would be a signature of
non-locality provided the detection takes place in a time shorter than it
would take a light pulse to reach $\tilde{z}$. However we can see that typical
cold neutron eigenstates in the gravitational field, which extend over a few
micrometers, would require time delays in the picosecond regime. Not only are
these delays minute in order to complete a measurement, but moreover this
would require the miror to move at a sizeable fraction of the speed of light.
Alternatively, the mirror should move on scales several orders of magnitude
below the micrometer regime, which is also unfeasible (and would lead to tiny,
probably undetectable effects anyway). The preparation of neutrons in extremely
high eigenstates (which would extend over large distances) does not appear to be realistic.

We can rescale the results displayed in Figs.\ \ref{fig-mak} and
\ref{fig-fixed} in order to reach regimes for which the moving mirror would
give rise to non-local effects. Indeed, reverting the numbers in these figures
to gravitational units (see Sec. \ref{bouncer}), we see that the light
velocity in gravitational units is given by $c(t_{g}/l_{g})=c\left(  4m/\hbar
g\right)  ^{1/3}$ and decreases for low masses and/or strong gravity fields.
For the parameters chosen in Figs.\ \ref{fig-mak} and \ref{fig-fixed}, a
measurement would have to be made in a time of the order of a gravitational
unit $t_{g}$, for an observer positoned at a few gravitational units $l_{g}$,
so the light velocity would need to be of the order $1\sim100$ in
gravitational units in order for measurements to take place outside the light cone
emanating from the mirror. This implies roughly that $g/m$ should be of the
order of $c^{3}/\hbar$, so that for the parameters displayed in
Figs.\ \ref{fig-mak} and \ref{fig-fixed} extremely low masses (compared to the neutron mass) and high gravity fields (compared to the gravitational field on the Earth's surface)
would lead to non-local effects.

\section{Conclusion}
\label{conc}
To sum up, we have investigated the effects produced by a moving mirror on the
dynamics of a quantum bouncer. We have seen, for different choices of
initial states, that a moving mirror modifies the properties of the quantum
bouncer, including for short times. Some effects are readily understandable from 
classical considerations, while other effects are purely quantum, coming 
form the fact that the wavefunction is modified everywhere
by the time-dependent boundary conditions. While in principle the moving
mirror could --at least formally -- give rise to non-local effects, we have
seen that such effects cannot be observed in the regimes accessed by current
experiments investigating the behavior of neutrons in a gravitational field.

\end{document}